\documentstyle[12pt,draft]{article}
 \newcommand \be {\begin{equation}}
\newcommand \bea {\begin{eqnarray} \nonumber }
\newcommand \ee {\end{equation}}
\newcommand \eea {\end{eqnarray}}

\newcommand{\bbi}{\noindent}

\begin{document}

\title{An introduction to the immune network
\\{\small Talk given at the EPS meeting, Florence 1993}}

\author{  Giorgio Parisi \\
Dipartimento di Fisica,
Universit\`a {\sl La  Sapienza}\\ INFN Sezione di Roma I \\
Piazzale
Aldo Moro, Roma 00187}
\maketitle

\begin{abstract}
	In this paper, after a telegraphic introduction to modern
immunology, we
present a simple model for the idiotypic network among
antibodies and we study
its relevance for the maintenance of  immunological memory.
We also consider the
problem of computing the memory capacity of such a model.

 \end{abstract}

\section{Introduction}

The aim of this talk is to give to the audience some feelings on
the
research that has been done in these recent years on the field
of the immune
network. Although I will concentrate my attention mainly on
the theoretical
aspects, it is necessary to recall some of the basic properties of
the immune
system, especially since I speaking to an audience of
physicists\footnote{
 A nice introduction to immunology is Hood et al. 1984.}.
This task is  no simple: immunology is a very much developed
science: 20.000
articles  appear every year\footnote{ Here I will provide only a
very short reference list, quoting only those
papers which may be easier to read.}.

In presenting this introduction, for simplicity I will do sharp
statements
that often should be better qualified. At the contrary of
physics, in biology
every law, included this one, has exceptions.

\section{Immunisation}

The most important function of the immune system, which
everybody
knows, is immunisation.  As one finds in the books, the
introduction of a given
amount of antigen (most of the antigens are proteins, e. g.
tetanus toxoid or
bovine albumin) inside a mammal\footnote{ The immune
system is present in all chordates (vertebrates), with some
difference among the different classes. For simplicity I will
refers only to
mammals.} stimulates the production of antibodies
directed against that given antigen.

We could just write the equation:
\be
Antigen\  =\  new \  protein.
\ee
Antibodies are soluble proteins secreted by the lymphocytes,
which have
a high affinity toward the antigen (in this context high affinity
means an
equilibrium constant K equal $10^5$ or more).

The precise number of chemically different antibodies, that an
organism
(e.g. a mouse) is able to produce at a given moment (i.e. the
repertoire), is of
order $10^6-10^7$. The number of antibodies is so high that
the repertoire of
possible responses is complete, i.e. the immune system is able
to react against
any possible protein.

 After the introduction of the antigen, only those lymphocytes
(B
lymphocytes), which have a high affinity with the antigen, are
stimulated and
they expand exponentially. This clonal expansion is crucial to
direct the
antibodies production against the antigen. The typical time,
that is needed to
reach the maximum of antibodies production, is of the order of
one week.

\section{ Some complications}

Things are no so simple as it may seem. The situation is
complicated by
the phenomenon of hypersomatic mutation. We have seen that
after the
introduction of the antigen there is competitive selection
among different
clones. However mutations may happen; in this way new clones
are created
and they are further selected. Normally the mutation rate of a
usual cell is
very low (of the order $10^{-9}$ per base per division),
however in this case we
have a much higher mutation rate (of the order $10^{-3}$). This
process leads to
variation of the immune response with time. Its study is an
interesting
problem in population dynamics. It is usually believe the
hypersomatic
mutation is present to further increase the affinity of the
antigens (Berek \&
Milstein 1988).

One should also say that the interaction with T lymphocytes is
crucial in
the dynamics of the system. These lymphocytes control the
activity of the B
lymphocytes and they may have an excitatory or a suppressor
role. Also other
cells (dendritic cells or macrophages), which present antigens
to B cells, are
crucial for the correct functioning of the system. We are going
to neglect all
these important details (and the tens of different kinds of
receptors on the
surface of the immune system cells) that would make this talk
quite long.

\section{ Tolerance}

A crucial effect of all these complicated interactions is that the
response
to the external antigen has a bell shaped form. To be more
precise let us
consider the following experiment.

We inject a dose x of antigen at time zero (priming) and at time
$t$ (e.g. 1
month later) we measure the production of antibody directed
against the
antigen shortly after the second shot. The response has
function of $x$ has a bell
shaped form. We can distinguish four regions:
\begin{itemize}
\item a) Very low $x$: no effect.
\item b) Low $x$: a decrease (not an increase!) of the antibody
production, i.e. low
dose tolerance.
\item c) Medium $x$: an increase of the antibody production,
which is the naively
expected result.
\item d) High $x$: a decrease (which may be very strong) of
the antibody
production, i.e. high dose tolerance. In other words a too strong
stimulation
leads to anergy.
\end{itemize}

Summarizing, for
both small and high dose of the antigen the responce after
priming  smaller than
without priming (tolerance). Only for medium dose of the
antigen the responce is
higher. The range of values for which a positive responce is
obtained depends on
the antigen. It is usually wide a few orders of magnitudo.

These unexpected features of the response are related to a
very  important
phenomenon, i.e. tolerance. In the nutshell it is crucial that the
immune system
does not react against the self, i.e. it does not produce
antibodies against his
own proteins. If this happens, severe illnesses (e.g. {\sl
diabetus mellitus}) may
arise.

As a consequence during the ontogenesis\footnote{
There is a large body of experimental evidence that the
immune system is
genetically able to produce antibodies against the self. The
learning happens
during the ontogenesis and it has not happened during the
phylogenesis.}  the immune
system learns not  to produce antibodies against the self.

Generally speaking we can conclude that when the immune
system is
stimulated, two pathways are open: tolerance or immunity; the
choice of the
pathway is crucial and depends on many factors.

A mechanism which contributes to the establishment of
tolerance is the
following. In order to be stimulated by B cells, T cells needs
two signal, one is
antigen specific and the other one is not specific\footnote{This
is the famous two signal theory (of
Bretcher and Cohn) and I am
particularly proud to have been the first to point out an
experimental proof of
its correctness (Parisi 1988).}. It was proved that in
presence of only the fist signal and in absence of the second
signal, T cells are
not stimulated and they are lead to anergy (Shwartz 1990).

\section{ The idiotypic network}

An antibody which starts to be produced at a certain moment
in large
quantities is (at the all practical effects) a new protein for the
organism. This
new antibody (which we call Ab1) stimulates the formation of
a second wave
of antibodies (Ab2) which are directed against Ab1. This fact is
not only
reasonable for a theoretically point of view, but it is
experimentally well
proved (Oudin \& Mitchel 1963).

In the same way it is reasonable to assume that Ab2 stimulates
the
formation of a third wave of antibodies Ab3 which are directed
against Ab2.
Not too much ingenuity is needed to assume that Ab3
stimulates the
production of Ab4 which stimulates....

In other words the production of a given antibody influence
the
production of other antibodies and we can thus speak of a
network of
antibodies (the so called idiotypic network, proposed by Jerne
1967, 1974,
1984).

A crucial property of the network is a symmetry of the
interaction: if Ab1
stimulates Ab2, Ab2 stimulates Ab1. As a consequence Ab3 is
not too different
from Ab1 and a large fraction of Ab3 binds to Ab2 and to the
antigen. Indeed
most of Ab2 looks (at the binding site) like the antigen.

The origin of this symmetry is not surprising. Protein - protein
interaction is dominated by weak short range non covalent
forces which
depends on the geometry (dipole dipole interaction), on the
charge distribution
and on hydrophilic - hydrophobic effects. In this situation is
clear that a strong
interaction is present only if the geometries of the two proteins
are
complementary (and there is match of the charge distributions
and of the
hydrophobicity).

While there are no doubts on the existence of the idiotypic
network, its
physiological relevance has been much debated (Cohn 1986,
Holberg et al.
1986, Hoffmann et al. 1988, Rossi et al. 1989).  It is been often
suggest that it
may be important for memory and this will be discussed in the
next section.

\section{ Memory}

As it should be clear from the previous discussion there are
two kinds of
memories:
\begin{itemize}
\item a) a positive one which is connected to immunisation.
\item b) a negative one, which is connected to tolerance,
especially self tole-
rance, i.e. the absence of antibodies directed against the
self\footnote{The precise
 definition of the self is a little ambigous: if antibodies belong
to
the self, Ab2 is an antibody directed against the self (Coutinho
1989). It should
also be stressed the total amount of antibody is constant as
function of time
and does not depend on immunisation (Varela et al. 1991).
Also in absence of
external antigens there is a large production of natural
antibodies. Their
physiological role is not completely clear.}.
\end{itemize}

The memory lasts for a very long time (practically infinite). If
we
concentrate our attention on the positive memory, the
explanations for this
phenomenon are of two different types.

\begin{itemize}
\item{\bf Static explanations}

	In this kind of explanations the dynamics and the
idiotypic network play
no role. Let me quote two of them:
\begin{itemize}
\item
a) Antibodies producing cells and their descendants have very
long mean
life.
\item
	b) The antigen remains for very long time in the
organism on the surface
of APC (antigen presenting cells, i.e. dendritic cells,
macrophages).
\end{itemize}
The mathematical equation describing situation (a) (in absence
of the
antigen) would be given in a first approximation by
\be
{dx_i \over dt}= 0,
\ee

where $x_i$ is the concentration of the $i$-th antigen. The
index $i$ runs from 0
to $N$, where $N$ is a large number (of the order of $10^8$ ).
The solution of this
equation is an easy task.

\item{\bf Dynamic explanations}

A typical explanation is the following. The network has two
stable state:
in the first the antibody Ab1 is not produced and in the second
the antibody
Ab1 is produced. The effect of the antigen is to induce the
transition of the
network from the first state to the second one. After the
transition the antigen
may disappear and memory lasts forever.

It is clear in this kind of explanation the number of stable
states should
be extremely large, i.e. at least of the order of all possible
combination of
antibodies which can be simultaneously produced (Parisi1989).

The mathematical equations describing this situation (in
absence of the
antigen) would be given in a first approximation by
\be
{dx_i \over dt}= -x_i + s_i[x],
\ee
where the functions $s_i[x]$, (which depends explicitly on the
index $i$ and
represent the stimulatory effects of the other antibodies) are
functions of all
the antibodies concentrations of the systems.

The stable states are the solutions of the equations:
\be
x_i = s_i[x],
		\label{TRE}
\ee
which has only the trivial solution ($x_i = 0$) if we neglect the
functions $s_i$.

As we have already remarked the functions si should be such
that the
equations \ref{TRE} must have an exponentially large number
of solutions.

\end{itemize}

If we do an analogy with electronics, the static explanation
corresponds to
dynamic memory chips: a capacitor is charged and it remains
active for a very
long time (a few milliseconds which is small on the human
scale but is large if
measured in nanoseconds.)

On the contrary the dynamic explanation corresponds to static
memory
chips: learning correspond to changing the state of a flip flop.
Here the
memory has an infinite mean life, i.e. it lasts as far as electric
power is on.

Unfortunately there is no consensus among immunologists on
which kind
of explanation should be correct. One can present arguments in
different
direction and no conclusive experiment has been done. It is
likely that in order
to achieve further progresses it is necessary do present a
detailed model and
to compare the predictions of the model with the experimental
data (Weisbuch
et al. 1990, Seiden \& Celada, 1992 Bern et al. 1993).

\section{ Models}

A large literature exists on the construction of models of the
immune
system. If we consider all the complications of the immune
system, a realistic
model cannot be simple.

Here due to limits of time I will consider a very simple model
(Parisi
1990), which should hopefully capture most of the qualitative
features of the
immune system and it has some points in common with the
Hopfield model
(Hopfield 1982).

We introduce a matrix $J_{i,k}$, which codes the effect of the
$k$-th antibody on
the ith antibody. The stimulatory effect of the network on the
$i$-th antibody is
given by
\be
h_i= \sum_k J_{i,k} x_k .
\ee

The equation of motion (in presence of the antigen
concentration $a_i$) are
\be
{dx_i \over dt} = f(x_i,h_i+a_i)			\label{CINQUE}
\ee
where f is given function.

If the matrix $J$ is symmetric\footnote{The symmetry of the
interaction has been
suggested in Cooper Willis \&
Hoffman (1983) and Rajewsky (1983).}, the equations are of
gradient type\footnote
{We also need $J_{i,i}=0$.}
and the  time evolution is such to bring the system toward a
fixed point of eq.
\ref{CINQUE}, i.e. no  chaotic behaviour is possible\footnote
{We have neglected the possibility the matrix $J$ is time
dependent. This may
arise from the variation (as function of time) in the populations
of produced
antibody, e.g. as an effect of hypersomatic mutation. The
relevance of a
possible time dependence in $J$ is difficult to estimate.}.

Different models depends on the choice of the function $f$ and
of the matrix
$J$. We can consider two different class of model, depending on
the connectivity
of the network (Perelson 1988 and Stewart \& Varela 1989).
\begin{itemize}
\item
a) Localised models: here the effect of a perturbation (i.e. a
variation of
$a_i$) is localised only on a few values of $k$. This may be
realised if for given i
there is a small number of $k$Õs for which $J_{i,k}$ is large and
the perturbation
does  not spread.
\item b) Percolating models: here the effect of a perturbation
influences the
whole system. There is a large number (e.g. proportional to
$N^{1/2}$) of $k$'s
which  are strongly affected by a perturbation at $i$. This may
be realised if for
given $i$  there is a large number of values of $k$'s for which
$J_{i,k}$ is large
or if there is a  small number of values of $k$'s for which
$J_{i,k}$ is large, but
the perturbation does  spread.
\end{itemize}

Both kinds of models can be constructed in such a way that the
number
of stable states is exponentially large (i.e. proportional to
$\exp(AN)$, with
positive $A$). An analytic computation of the number of fixed
point can be done
using the replica tecnique (M\'ezard et al. 1987) in the case of
randomly chosen
$J$. Learning however will be quite model depending and we
cannot discuss it in
details (Lefevre \& Parisi 1992).

\section{ Conclusions}

The organisation of the immunological system as a network is
not a
isolated phenomenon in biology (Cattaneo 1991). In many
other cases
networks of interacting substances are present (e.g.
intracellular signaling,
hormones in an organism). The crucial problem is to
understand which
functions of the immune system crucially depend on the
network.
At the present moment there are many important problems in
the
immune system that are not well understood (e.g. the very
existence and
structure of suppressor T cells). The construction of a realistic
model is very
difficult due to these incertitudes.

New experimental data must be collected in order to test the
network
hypothesis. For example the experimental study of the time
dependence of the
level of natural antibodies (i.e. antibodies produced in absence
of external
antigens) show the presence of oscillations and chaotic
behaviour. These
oscillations strongly suggest the presence of underlying non
linear evolution
equations (Varela et al. 1992).

A particular field in which a better understanding of the
network would
be extremely useful is related to tolerance. It is rather difficult
to break
tolerance or to induce tolerance. If tolerance could be induced
at our will, we
could cure autoimmune diseases (like {\sl diabetus mellitus})
or perform organs
transplantation without rejection. If tolerance is (at least
partly) under the
control of the network, a better understanding of the network
may strongly
improve our ability in dealing with those problems.

\section*{References}

 Berek C. \& Milstein C. (1988), Immunol. Rev. 105, 5.

\bbi Bern U., van Hemmen J. L. \& Sulzer B., J. (1993), Theor.
Biol. 165, 1.

\bbi Cattaneo A. (1991), in {\sl Neural networks: from biology
to high energy
physycs},  ed. by Benhar O., Bosio C., Del Giudice P. and Tabet E.,
(ETS editrice
Pisa).

\bbi Cohn M. (1986),  Ann. Immunol. (Inst. Pasteur) 137C, 64-
76.

\bbi Cooper-Willis A. \& Hoffmann G. W. (1983), Mol. Immunol.
20, 865.

\bbi Couthinho A. (1989), Immunol. Rev. 110, 63.

\bbi Hoffmann G. W., Kion T. A., Forsyth R. B., Soga K. G. and
Cooper-Willis (1988), in
{\sl Theoretical Immunology}, ed. by A. Perelson, Addison-
Wesley, New
Jersey.

\bbi Holberg D., Freitas A., Portnoi D., Jacquemart F., Avrameas
S. \& Coutinho A.
(1986), Immun. Rev. 93, 147-178.

\bbi Hood L.E., Weissman I. L., Wood W.B. \& Wilson J.H.
(1984), {\sl Immunology},
(Benjamin Pu. Co., Menlo Park).

\bbi Hopfield J. J. (1982), Proc. Nat. Acad. Sci. (USA) 79, 2554-
2558.

\bbi Jerne N. K. (1967), in {\sl The Neurosciences: A Study
Program, eds. Quarton,
G.,  Melnechuk, T \& Schimitt, F. O.} (The Rockfeller Univ. Press
New York), pp.
200-208.

\bbi Jerne N. K. (1974), Ann. Immunol. (Inst. Pasteur) 125C,
1127-1137.

\bbi Jerne N. K. (1984), Immunol. Rev. 79, 5-24.

\bbi Lefevre O. and Parisi G. (1992), Network 4, 39.

\bbi M\`ezard M., Parisi G. \& Virasoro M. (1987), {\sl Spin
glass theory and
beyond}, Word  Scientific, Singapore.

\bbi Oudin J \& Michel M. (1963), C. R. Seances Acad. Sci. 257,
805-610.

\bbi Parisi G. (1988), Ann. Immunol. (Inst. Pasteur) 139, 177.

\bbi Parisi G. (1989), Phys. Rep. 184, 283.

\bbi Parisi G. (1990), Proc. Natl. Acad. Sci. 87, 429.

\bbi Perelson A. (1988) {\sl Theoretical Immunology, SFI
Series on the Science of
Complexity}, Addison Wesley, New Jersey.

\bbi Rajewsky K. (1983), Ann. Immunol. (Inst. Pasteur) 134D,
133.

\bbi Rossi F., Dietrich \& Kazachzine M. (1989), Immunol. Rev.
110, 135.

\bbi Seiden P.E. \& Celada F., J. theor. Biol (1992) 158, 329.

\bbi Stewart J. \& Varela F. J. (1989), Immunol. Rev. 110, 37.

\bbi Shwartz (1990), Science  248, 374.

\bbi Varela F. J., Anderson A., Dietrich G. Sumblad A., Holmberg
D., Kazachzine M. \&
Coutinho A. (1991), Proc. Natl. Acad. Sci. 88, 5917.

\bbi Weisbuch, G., De Boer R. and Perelson A. S. (1990), J.
Theoret. Biol. 146, 483.

\end{document}